\documentclass[10pt,oneside,twocolumn,a4paper]{article}
\usepackage{graphicx}
\usepackage[utf8]{inputenc}
\usepackage{mathptmx}
\usepackage{tabularx}
\usepackage{ragged2e}
\usepackage[singlelinecheck=false]{caption}
\usepackage[T1]{fontenc}
\usepackage[numbers]{natbib}
\usepackage{amsmath}
\usepackage{siunitx}
\usepackage{algorithm}
\usepackage{balance}
\usepackage[]{algpseudocode} 
\usepackage[table,xcdraw]{xcolor}
\usepackage[acronym,nonumberlist,nopostdot,nogroupskip]{glossaries}
\PassOptionsToPackage{hyphens}{url}\usepackage[hidelinks]{hyperref}
\RequirePackage[blocks]{authblk}
\usepackage[english]{babel}
\makeatletter
\let\ps@plain\ps@empty
\def\@xivpt{14bp}

\def\@sect#1#2#3#4#5#6[#7]#8{%
  \ifnum #2>\c@secnumdepth
    \let\@svsec\@empty
  \else
    \refstepcounter{#1}%
    \protected@edef\@svsec{%
      \ifnum #2<4
        \hb@xt@10mm{\csname the#1\endcsname}\relax
      \else
        \hb@xt@12mm{\csname the#1\endcsname}\relax
      \fi}%
  \fi
  \@tempskipa #5\relax
  \ifdim \@tempskipa>\z@
    \begingroup
      #6{%
        \@hangfrom{\hskip #3\relax\@svsec}%
          \interlinepenalty \@M #8\@@par}%
    \endgroup
    \csname #1mark\endcsname{#7}%
    \addcontentsline{toc}{#1}{%
      \ifnum #2>\c@secnumdepth \else
        \protect\numberline{\csname the#1\endcsname}%
      \fi
      #7}%
  \else
    \def\@svsechd{%
      #6{\hskip #3\relax
      \@svsec #8}%
      \csname #1mark\endcsname{#7}%
      \addcontentsline{toc}{#1}{%
        \ifnum #2>\c@secnumdepth \else
          \protect\numberline{\csname the#1\endcsname}%
        \fi
        #7}}%
  \fi
  \@xsect{#5}}
\renewcommand\LARGE{\@setfontsize\LARGE{16}{20}}
\def\abstract#1{\def\@abstract{#1}}
\def\abstractEn#1{\def\@abstractEn{#1}}
\def\titleEn#1{\def\@titleEn{#1}}
\def\@maketitle{%
  \newpage
  \null
  \let \footnote \thanks
    {\LARGE\bfseries\RaggedRight \@title \par}%
    \vskip 1\baselineskip%
    {\normalsize
      \@author\par}%
    \vskip \baselineskip%
    {\section*{Abstract}
      \@abstractEn}%
  \par
  \vskip 3\baselineskip}

\renewcommand\section{\@startsection {section}{1}{\z@}%
                                   {-3.5ex \@plus -1ex \@minus -.2ex}%
                                   {\baselineskip}%
                                   {\normalfont\Large\bfseries\RaggedRight}}
\renewcommand\subsection{\@startsection{subsection}{2}{\z@}%
                                     {\baselineskip}%
                                     {1ex}%
                                     {\normalfont\large\bfseries\RaggedRight}}
\renewcommand\subsubsection{\@startsection{subsubsection}{3}{\z@}%
                                     {1\baselineskip}%
                                     {3bp}%
                                     {\normalfont\normalsize\bfseries\RaggedRight}}
\renewcommand\paragraph{\@startsection{paragraph}{4}{\z@}%
                                    {1\baselineskip\@plus1ex \@minus.2ex}%
                                    {3bp}%
                                    {\normalfont\normalsize\RaggedRight}}
\renewcommand\subparagraph{\@startsection{subparagraph}{5}{\parindent}%
                                       {3.25ex \@plus1ex \@minus .2ex}%
                                       {-1em}%
                                      {\normalfont\normalsize\bfseries\RaggedRight}}

\usepackage{listings}
\definecolor{darkred}{rgb}{0.6,0.0,0.0}
\definecolor{darkgreen}{rgb}{0,0.50,0}
\definecolor{lightblue}{rgb}{0.0,0.42,0.91}
\definecolor{orange}{rgb}{0.99,0.48,0.13}
\definecolor{grass}{rgb}{0.18,0.80,0.18}
\definecolor{pink}{rgb}{0.97,0.15,0.45}

\lstdefinestyle{colored}{ %
  basicstyle=\scriptsize\ttfamily,
  backgroundcolor=\color{white},
  commentstyle=\color{darkgreen}\itshape,
  keywordstyle=\color{blue}\bfseries,
  stringstyle=\color{darkred},
}

\makeatother
\DeclareCaptionLabelSeparator{enskip}{\enskip}
\newcolumntype{Z}[1]{%
  >{\arraybackslash}p{#1}%
}
\newcolumntype{Y}{>{\centering\arraybackslash}X}

\usepackage[format=plain,labelsep=period,justification=justified,font=small]{caption}

\title{A Configurable and Efficient Memory Hierarchy for Neural Network Hardware Accelerator
\thanks{This work was supported in part by the German Federal Ministry of Education and
Research (BMBF) within the projects Scale4Edge under contract no. 16ME0129 and MANNHEIM-FlexKI under contract no. 01IS22086H.}}
\renewcommand\footnotemark{}

\author{Oliver Bause}
\author{Paul Palomero Bernardo}
\author{Oliver Bringmann}
\affil{University of Tübingen}
\affil{Chair for Embedded Systems}
\affil{oliver.bause@uni-tuebingen.de, paul.palomero-bernardo@uni-tuebingen.de, oliver.bringmann@uni-tuebingen.de}

\loadglsentries{glossary.sty}

\abstractEn{As machine learning applications continue to evolve, the demand for efficient hardware accelerators, specifically tailored for deep neural networks (DNNs), becomes increasingly vital. In this paper, we propose a configurable memory hierarchy framework tailored for per layer adaptive memory access patterns of DNNs. The hierarchy requests data on-demand from the off-chip memory to provide it to the accelerator's compute units. The objective is to strike an optimized balance between minimizing the required memory capacity and maintaining high accelerator performance. The framework is characterized by its configurability, allowing the creation of a tailored memory hierarchy with up to five levels. Furthermore, the framework incorporates an optional shift register as final level to increase the flexibility of the memory management process. A comprehensive loop-nest analysis of DNN layers shows that the framework can efficiently execute the access patterns of most loop unrolls. Synthesis results and a case study of the DNN accelerator UltraTrail indicate a possible reduction in chip area of up to $62.2\%$ as smaller memory modules can be used. At the same time, the performance loss can be minimized to $2.4\%$.} 

\begin{document}

\maketitle

\section{Introduction}
In recent years, machine learning has revolutionized various fields including video, sound, and language processing. \Glspl{dnn} can learn and recognize complex patterns in data by building a hierarchy of increasingly complex features, starting from simple edge detection and progressing to more abstract concepts like object recognition \cite{goodfellow2016deep}. In embedded systems, hardware accelerators can greatly improve the required runtime of \glspl{dnn}. These accelerators are specialized hardware modules designed to speed up the highly parallel mathematical operations of running neural networks. This can be particularly useful in areas where resources are limited and the ability to run \glspl{dnn} efficiently is crucial. The \gls{cpu} of a \gls{mc} cannot efficiently perform the processing of these data-parallel networks as they utilize only a few cores. Additionally, their broad instruction set will not be used since neural networks often require basic arithmetic operations. Therefore, a growing number of specialized hardware accelerators are being developed \cite{howard2017mobilenets, bernardo2020ultratrail, kyriakos2019high, ajani2021overview}. Requirements for accelerators include small chip size, low power consumption, and implementation of a memory interface that minimizes the number of expensive accesses to the off-chip memory \cite{siu2018memory}.

The development of a suitable on-chip memory hierarchy is particularly important because the required memory occupies a large portion of the total chip area and is responsible for a major part of the accelerator’s energy consumption. Thus, a trade-off must be found toward determining the minimal capacity required to maximize the accelerator’s performance and throughput. Developing and evaluating different designs is time-consuming. The sheer size of the design space itself renders it practically impossible to manually test all possible configurations, which can result in potentially superior memory architectures being passed over. Thus, assisting the engineer to explore and evaluate a larger portion of the available design space more easily and quickly could improve the implemented architecture and simultaneously reduce development time.

In this paper, we introduce a configurable memory framework that can semi-automatically generate and test an efficient memory hierarchy with multiple hierarchy levels and banks, different memory macros, and a pattern-based prefetching algorithm. Loop analyses of neural networks are used to determine their memory accesses and create possible framework configurations. The resulting simulation and synthesis reports can be used by engineers to select the most suitable memory hierarchy for their application.

\section{Related Work}
In the area of neural network accelerator design, the optimization of memory structures is a critical consideration, given their substantial impact on both area and power requirements. To support designers, \gls{dse} tools were developed. They utilize high level system models to evaluate different design options, like different memory configurations or type and amount of processing units. The automated search of the enormous available design space can accelerate the development time while simultaneously improving the final design in regard of chip area, power consumption, and production costs \cite{pimentel2016exploring}. The \gls{dse} tool ZigZag \cite{mei2020zigzag, mei2021zigzag} stands out for its systematic approach to exploring the design space, with a specific emphasis on memory architectures. By enabling efficient evaluation of diverse memory configurations, it facilitates the identification of optimal choices that strike a delicate balance between performance and resource utilization.
However, ZigZag does not provide a hardware implementation of the found memory hierarchies, which is a crucial step when aiming for accurate performance, power, and area measures and the ability for fast prototyping in existing \gls{rtl} designs.

Complementary to ZigZag, other works contribute valuable insights into memory optimization techniques tailored to the specific requirements of neural network workloads. For instance, \cite{howard2017mobilenets}, \cite{li2016optimizing}, and \cite{li2018optimizing} delve into strategies such as weight pruning, quantization, memory compression, and minimizing off-chip memory accesses. These techniques aim to reduce the memory footprint while preserving the accuracy of neural network models, thereby addressing the challenges posed by limited resources in hardware accelerators.

\section{Embedded Memories}
Memory has always been a major aspect of the development of computing systems. However, especially the performance of non-volatile memories is usually several orders of magnitude slower than that of the \glspl{dpu}, thus increasing the dependence of the overall system performance on memory-access optimization. As a result, the most efficient utilization of memory should always be taken into account in the development of applications and hardware, respectively.

Memory architectures can easily take up more than 60\% of the entire \gls{soc} chip area. Therefore, they are also responsible for a huge fraction of the total power consumption. Embedded systems mostly have strict requirements in terms of size, power, and cost. Thus, the goal is to reduce production costs and avoid heat generation which could result in performance deterioration, possible hardware damage, and shortened battery life \cite{macii2002memory, rajsuman2001design}.

\Glspl{sram} play a crucial role in the proposed memory hierarchy framework, serving as integral components in the memory architecture. \Glspl{sram}, characterized by their ability to retain stored information without the need for periodic refreshing, contribute to the framework's efficiency and speed. Their utilization in the memory hierarchy enables rapid access and retrieval of data within one clock cycle \cite{gunasekaran_2019}.

\subsection{Memory Requirements}
A \gls{dnn} has multiple types of data to store: features, weights, and intermediate results. The maximum and current size of these data types highly depends on the architectural design and data flow of the neural network. The resulting storage requirements of the features and weights from common neural networks like \gls{tcresnet} \cite{choi2019temporal} and AlexNet \cite{krizhevsky2017imagenet} can range from only \SI{64}{\kilo\byte} to more than \SI{500}{\mega\byte}, respectively. Therefore, \gls{dnn} hardware accelerators often utilize small on-chip memories and only load the currently needed subset of data from a larger off-chip memory. However, these off-chip memory accesses are time-consuming and require up to two orders of magnitude more energy than the on-chip memory accesses. 

Siu et al. \cite{siu2018memory} concluded, that the peak memory requirements are often dominated by only a few convolutional layers.
As a result, it may be reasonable to tailor the on-chip memory architecture towards the average memory requirements.
The consequent loss in performance can be justified by the substantial reduction in chip area and energy requirements.

\subsection{Memory Access Pattern}
\begin{figure}[t]
	\centering
	\includegraphics[width=\linewidth]{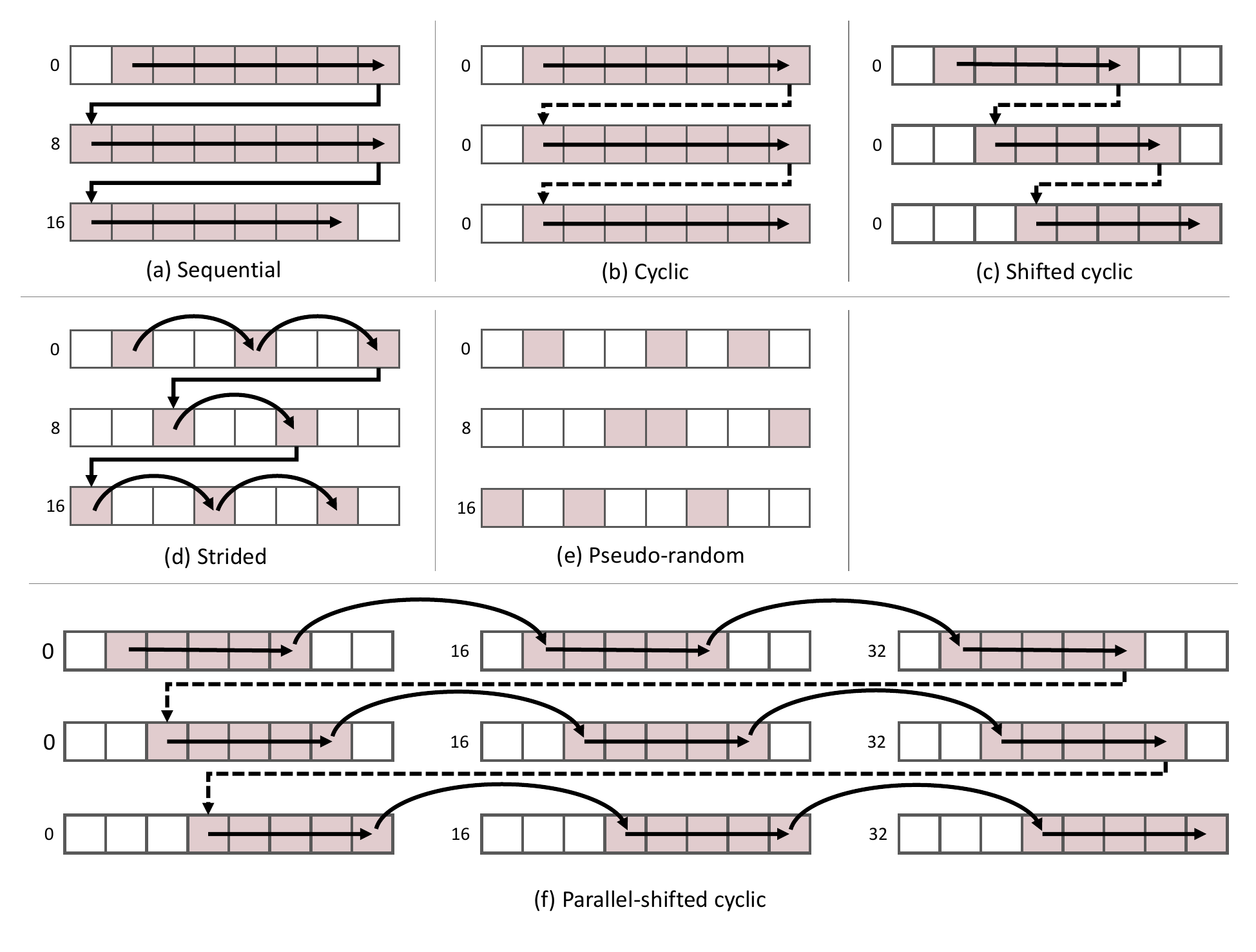}
	\caption[Schematic presentation of the five memory access patterns]{Schematic view of the memory access patterns (a) sequential, (b) cyclic, (c) shifted cyclic, (d) strided, (e) pseudo-random, and (f) parallel-shifted cyclic. The number left to a memory bank is its start address. Note that all banks of the cyclic patterns start with \textit{0} to highlight that they access the same addresses multiple times. This figure is inspired by \cite{jang2010exploiting}.}
  \label{fig:access_pattern}
\end{figure}
\Glspl{dnn} are commonly trained and executed on parallel architectures like 2D systolic arrays, \gls{simd}, or \gls{mimd} compute units. Each approach has its benefits and drawbacks. 2D systolic arrays, as an example, are dedicated \glspl{dpu} performing static functions, enabling them to require very little chip area. However, the array offers almost no flexibility. The instruction cannot be changed after the synthesis and individual \glspl{dpu} cannot be stalled \cite{das_krishna_2018}.

Nevertheless, all architectures need several data words for each clock cycle, which have to be loaded from memory beforehand. Depending on the \gls{dnn}'s data flow and layer mapping onto the target hardware, different memory access patterns can occur. This paper will focus on the patterns shown in Figure \ref{fig:access_pattern}, that are classified as inspired by Byunghyun et al.'s research \cite{jang2010exploiting}.
\begin{itemize}
	\item[a)] \textit{Sequential}: The data is stored in the memory on successive addresses and accessed sequentially starting from a base address. Every address is accessed exactly once, whereby there is no data reuse.
	\item[b)] \textit{Cyclic}: The cyclic pattern has a defined cycle length $l$. Like the \textit{sequential} pattern, data is accessed at successive addresses in the memory, but after loading $l$ data words, the pattern returns to the base address and starts again. Therefore, the pattern accesses a constant data stream and if the memory architecture can buffer the whole cycle, the data from every off-chip memory address will only be read once.
	\item[c)] \textit{Shifted cyclic} or \textit{overlapping}: The shifted cyclic pattern is an extension of the \textit{cyclic} pattern with the addition that, after the completion of each cycle, the base address will be shifted by some constant $s$. As a result, the individual cycles are overlapping and a portion of the required data has already been loaded in a previous cycle, the rest has to be preloaded.
	\item[d)] \textit{Strided}: In contrast to the previous patterns, data in a strided pattern is not stored at successive memory addresses, but have a constant offset from each other. The data of addresses in between is not requested and thus skipped. Note that it is possible to combine this pattern with the (shifted) cyclic pattern.
	\item[e)] \textit{Pseudo-random}: In the pseudo-random access pattern, non pre-calculable addresses are accessed. A determination of the next targeted addresses is hardly possible, if at all.
    \item[f)] \textit{Parallel-shifted cyclic}: The access pattern can also consist of multiple simple patterns, that are executed in parallel or are nested within each other. The parallel-shifted cyclic pattern, for example, involves several shifted cyclic patterns that switches to the next pattern after one cycle. Only after all patterns have run through one cycle each, the outer pattern returns to the first shifted-cyclic pattern and executes the shift of each nested pattern.
\end{itemize}

\section{Design Concept of the Memory Framework}
The design concept of the memory framework was derived from the goal to minimize the memory footprint of an accelerator's memory modules. The memory requirements regarding power consumption, chip area, capacity, and throughput are tightly coupled to the data flow through the accelerator and its use-case. A high data reuse percentage decreases the number of expensive off-chip memory accesses. However, depending on the data access pattern, the needed storage requirements to achieve a higher data reuse rate can grow significantly. As a result, the chip area and energy consumption will increase. Therefore, an optimal co-design between the network’s data flow and
available hardware configurations must be found.

The design space of hardware accelerators, however, has reached an extent, that is no longer manually explorable. Engineers have a large degree of freedom in adjusting data flow and selecting the right hardware components so that \gls{dse} tools are needed to support and accelerate the design process. However, they mostly only utilize single \gls{sram} or \gls{dram} modules, but the integration of a memory hierarchy could increase performance, lower chip area and energy consumption, or a trade-off between both \cite{mei2020zigzag, das_krishna_2018}.

\subsection{Architecture}
The memory framework offers a highly parameterized and runtime-adjustable SystemVerilog template.
It is specialized for the deployment within neural network hardware accelerators and could be integrated into existing \gls{dse} tools.
The parameters include:
\begin{itemize}
    \item \textit{Off-chip interface}: The off-chip data width is needed to initiate the data input port. Additionally, the address width will be used to generate valid off-chip read requests.
    \item \textit{Hierarchy depth}: The number of generated hierarchy levels can range from one to five.
    \item \textit{Hierarchy level configuration}: Each hierarchy level is configured independently. The settings for each level include the memory macro name, the number of banks, the word width and RAM depth of the given macro, and whether the memory module is single- or dual-ported.
    \item \textit{\Gls{osr}}: If the \gls{osr} is enabled, the bit width of the shift register and a list of available shifts have to be defined.
\end{itemize}

\begin{figure}[t]
	\centering
	\includegraphics[width=\linewidth]{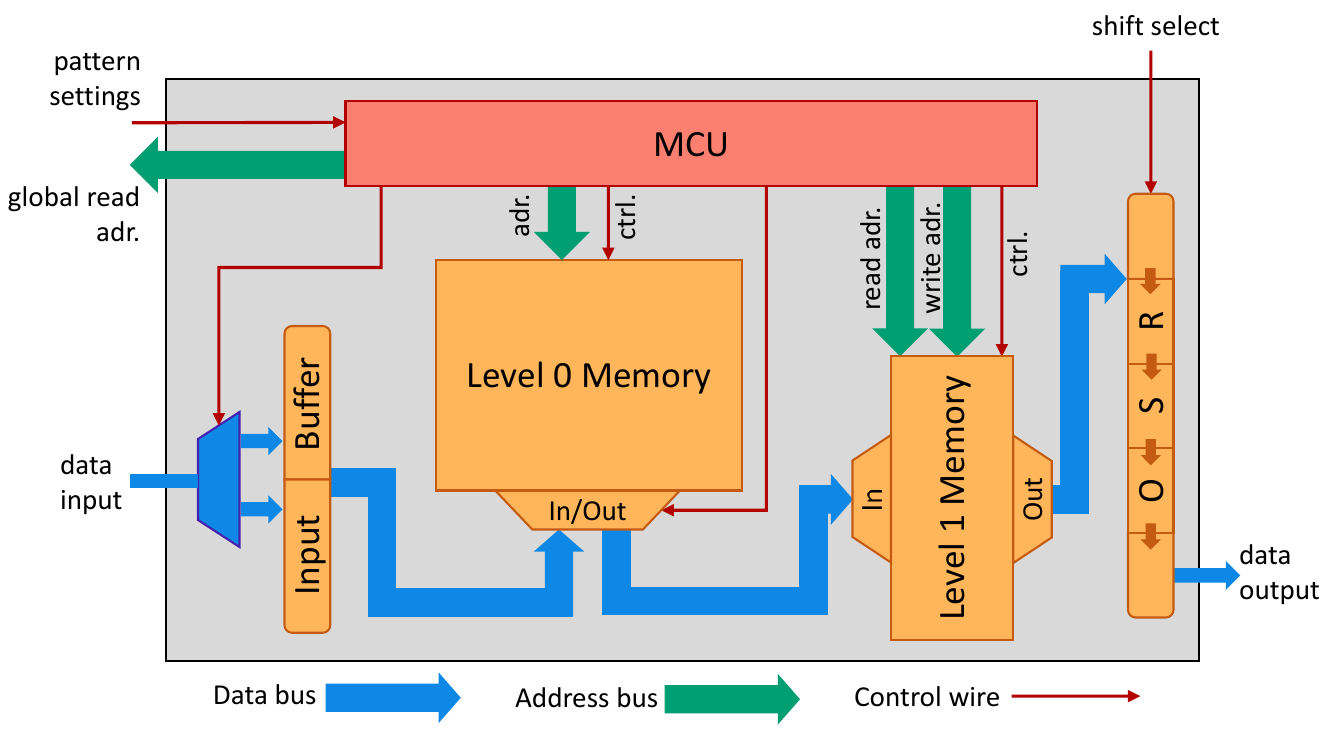}
	\caption{Overview of the memory framework's architectural design. This configuration is equipped with the optional \gls{osr} and two hierarchy levels where level 0 is a single-ported module with a higher capacity than the dual-ported memory of level 1. Note that level 1 needs two address buses (adr.), one for each port.}
  \label{fig:architecture}
\end{figure}
Figure \ref{fig:architecture} illustrates an exemplary memory framework configuration with two hierarchy levels. All memory modules are fully controlled by the on-chip \gls{mcu}.
The controller activates memory modules, adjusts read and write addresses of each module according to the currently selected access pattern and, most importantly, ensures data integrity and correctness within the whole hierarchy.
To preload data from the off-chip memory, an interface will request data from the global address space and buffer the incoming data stream.
The input buffer and interface are also controlled by the \gls{mcu}.
To support even more access patterns efficiently, the memory framework has an optional powerful \acrfull{osr}.
It is a register file located between the last hierarchy level and the accelerator's processing units.
The \gls{osr} buffers data one last time and performs customizable, non-unit shifts.
The memory hierarchy can be controlled fully by an off-chip \gls{mc} or partly by the accelerator.

\subsubsection{Input Buffer}
The data words received from the off-chip memory undergo temporary buffering in the input buffer before being stored in the memory hierarchy.
This addresses three key challenges.
The primary challenge arises from potential clock mismatches between the accelerator and the connected \gls{mc} \cite{bernardo2020ultratrail,lerch2021design}.
The input buffer is clocked by the \gls{mc}'s clock source.
This ensures synchronization between the input buffer and the off-chip \gls{mc} and prevents communication failures.
A handshake protocol is utilized to enable data transfer from the input buffer to the first hierarchy level.

Second, the input buffer aligns the incoming data stream for the hierarchy.
The buffer is a register file with the same word width as the first level of the hierarchy.
If the off-chip architecture has a smaller word width, the input buffer will hold multiple words before passing them to the hierarchy.
This optimizes memory efficiency and reduces the write load on the first level.

Lastly, the input buffer prevents potential blocking of the off-chip memory during data storage in the hierarchy, enhancing overall system performance by allowing the \gls{mc} to manage other tasks.
This dynamic buffering process ensures a smooth and synchronized flow of data, promoting stable communication between different clock domains and optimizing the use of memory resources.

\subsubsection{Memory Hierarchy}
The memory hierarchy, storing requested data words from the off-chip memory, is fully configurable with one to five levels. Each level can utilize different memory modules, macros, and types. In Figure \ref{fig:architecture}, level 0 is closest to the off-chip memory, forming the initial data pass-through point, while each subsequent level follows in the data flow. This nomenclature, contrary to traditional \gls{cpu} cache architectures, is a consequence of the data-flow-driven design. Unlike standard \gls{cpu} designs, hardware accelerators' predetermined data accesses render traditional caching strategies obsolete.

Each level of the hierarchy can feature one or two banks for data storage.
Dual-banked levels with single-ported memory modules strike a balance between performance and chip area.
However, both banks must utilize the same memory module.
The framework optimizes the virtual address space to evenly distribute the load between banks, minimizing the risk of simultaneous read and write attempts to the same single-ported bank.
In this use case, two single-ported memory banks are used to emulate a single dual-ported memory module.
Therefore, it is not reasonable to use more than two banks in a hierarchy level.
A maximum of one word can be written and read simultaneously per level in one clock cycle.
A more dynamic implementation, allowing multiple simultaneous reads and writes, would require significantly more chip area and power, which would defeat the purpose of the framework.

Additionally, to prevent system failures, the framework ensures that dual-ported memory modules do not access the same address with both ports simultaneously. If multiple levels are configured, all data must traverse each level, simplifying control mechanisms at the expense of a minor initialization phase. Note, that higher levels do not retain subsets of data from lower levels. They instantly clear memory space after the last specified pattern read was executed to expedite loading new data.

\subsubsection{Memory Control Unit}
The \gls{mcu} controls the framework from the data retrieval from the off-chip memory to the output of the last memory hierarchy level.
Furthermore, it calculates memory accesses of the selected pattern and ensures data integrity throughout the hierarchy’s runtime.

\begin{figure}[t]
	\centering
	\includegraphics[width=\linewidth]{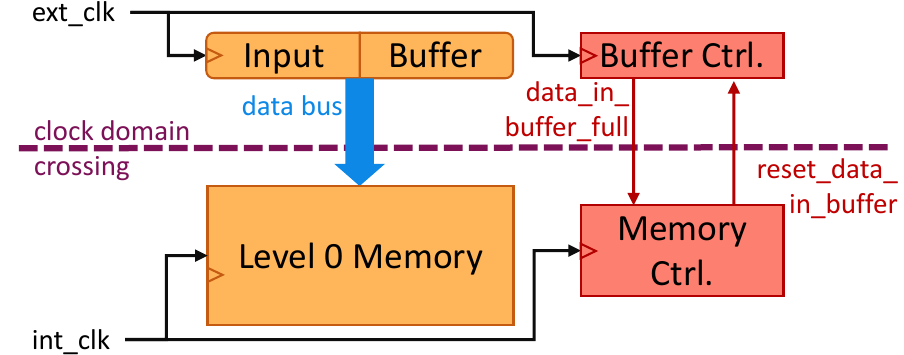}
	\caption[Clock domain crossing between input buffer and hierarchy level 0]{Clock domain crossing between the input buffer and the first hierarchy level. The input buffer is clocked by the faster external clock source while the memory hierarchy is clocked by the slower on-chip clock. To synchronize the data transfer, the two control wires, \textit{buffer full} and \textit{reset buffer}, are required.}
  \label{fig:cdc}
\end{figure}
The accelerator's clock source may operate at a different frequency than the external clock source, necessitating the implementation of synchronizers for bidirectional communication across the \gls{cdc}.
The \textit{buffer full} control signal and the data bus from the input buffer to the first hierarchy level must be synchronized to the accelerator's clock domain.
This involves holding the signal for at least an entire cycle to ensure stability in the slower domain and preventing metastability.
Metastability refers to a transient state where the output of a circuit is undetermined, leading to potential glitches and unpredictable behavior.
Additionally, challenges arise from the uncertainty regarding the number of cycles needed by the hierarchy to store its content, stemming from conflicts in bank access and RAM availability.
Figure \ref{fig:cdc} presents the handshake mechanism to synchronize communication between both clock domains.
The \textit{buffer full} signals the memory controller, that data is available to be written into the first hierarchy level.
After the write is complete, the memory controller sends a reset signal to the buffer controller.

\subsubsection{Pattern Selection and Calculation}
\begin{table}[t]
\resizebox{0.49\textwidth}{!}{%
\begin{tabular}{llZ{19em}}
\hline
\rowcolor[HTML]{C0C0C0} 
\textbf{Port} & \textbf{Scope} & \textbf{Description} \\ \hline
\rowcolor[HTML]{FFFFFF} 
\begin{tabular}[c]{@{}l@{}}internal\_\\ clk\_i\end{tabular} & hier. & Accelerator's clock source. \\
\rowcolor[HTML]{EFEFEF} 
\begin{tabular}[c]{@{}l@{}}external\_\\ clk\_i\end{tabular} & hier. & Off-chip clock source. \\
\rowcolor[HTML]{FFFFFF} 
reset\_i & hier. & Reset signal. \\
\rowcolor[HTML]{EFEFEF} 
data\_in\_i & hier. & Data input bus from off-chip memory modules. \\
\rowcolor[HTML]{FFFFFF} 
\begin{tabular}[c]{@{}l@{}}start\_\\ address\_i\end{tabular} & hier. & Relative or absolute address within the off-chip address space from which the framework will start to request data from. \\
\rowcolor[HTML]{EFEFEF} 
\begin{tabular}[c]{@{}l@{}}cycle\_\\ length\_i\end{tabular} & level & Pattern cycle length of each level. \\
\rowcolor[HTML]{FFFFFF} 
\begin{tabular}[c]{@{}l@{}}inter\_cycle\_\\ shift\_i\end{tabular} & level & Defines the number of data words the pattern cycle will be shifted after each completed cycle. An inter-cycle shift of $0$ results in a \textit{cyclic} pattern. If the inter-cycle shift is equal to the cycle length, the pattern will be \textit{linear}. \\
\rowcolor[HTML]{EFEFEF} 
skip\_shift\_i & level & Number of cycles that will be run before the inter-cycle shift is executed. \\
\rowcolor[HTML]{FFFFFF} 
\begin{tabular}[c]{@{}l@{}}disable\_\\ output\_i\end{tabular} & hier. & Disables the data output to the accelerator. The hierarchy will still preload data from the off-chip memory, if possible. \\
\rowcolor[HTML]{EFEFEF} 
\begin{tabular}[c]{@{}l@{}}shift\_\\ select\_i\end{tabular} & hier. & If the OSR is configured, the shift select is used to select the shift from the list of available shifts. 0 disables data output from the OSR. \\
\rowcolor[HTML]{FFFFFF} 
\begin{tabular}[c]{@{}l@{}}global\_read\_\\ address\_o\end{tabular} & hier. & Signals the off-chip memory address the framework requests to read from. \\
\rowcolor[HTML]{EFEFEF} 
data\_out\_o & hier. & Data output bus toward the accelerator. \\ \hline
\end{tabular}
}
\caption{List of the memory framework's ports. The scope indicates whether the port is unique for the whole framework or is an array with one entry for each hierarchy level configured.}
\label{tab:ports}
\end{table}

The \gls{mcu} executes the currently selected pattern which can be altered during a reset cycle.
The ports are described in Table \ref{tab:ports}.
The pattern management within the \gls{mcu} facilitates the creation of diverse cyclic, overlapping, and sequential patterns.
The starting point in the off-chip address space from which the memory hierarchy will request data from, can be configured by either an absolute or relative address.
The cycle length signifies the pattern iteration's length, where accessing the \textit{cycle length}-th element leads to a cyclical return to the first one.
To augment pattern flexibility, the inter-cycle shift introduces a customizable shift activated after \textit{skip shift} cycle completions.
However, the framework lacks runtime input validation, entrusting the engineer with the responsibility of understanding and aligning chosen parameters with the accelerator's architecture.
This design choice, avoiding runtime verification, is a deliberate trade-off to prevent unnecessary increases in chip area and power consumption, emphasizing the need for careful parameter selection and alignment with the overall architecture.

\begin{figure}[t]
	\centering
	\begin{lstlisting}[language=Verilog, mathescape=true, tabsize=2, style=colored, caption={Simplified \gls{mcu} pattern calculation executed in each hierarchy level.},label={code:mcu}]
// Write and read cycles of each hier. level
if write_enable[l] begin
	writing_pointer[l] = (writing_pointer[l] + 1) % RAM_DEPTH[l];
	data_reload_counter[l] -= 1;
end

write_enable = false;
if data_reload_counter[l] > 0 begin
    // Activate write cycle if RAM address is empty 
    // and previous level 
	if RAM[l][writing_pointer[l]] is empty and read_enable[l-1] begin
		write_enable[l] = true;
  end
end

// If the succeeding level was in a write cycle, read the next word
if write_enable[l+1] begin
	pattern_pointer[l] += 1;
	// If one pattern cycle is complete, perform inter-cycle shift
	if pattern_pointer[l] = cycle_length[l] begin
		pattern_pointer[l] = 0;
		skips[l] += 1;
		if skips[l] > skip_shifts[l] begin
			skips[l] = 0;
			offset_pointer[l] = (offset_pointer[l] + inter_cycle_shift[l]) % RAM_DEPTH[l];
			data_reload_counter[l] += offset_pointer;
    end
  end
end

read_pointer = (offset_pointer[l] + pattern_pointer[l]) % RAM_DEPTH[l];
if RAM[l][read_pointer] not empty begin
	read_enable[l] = true;
end
    \end{lstlisting}
\end{figure}

\begin{figure}[t]
    \centering
	\includegraphics[width=\linewidth]{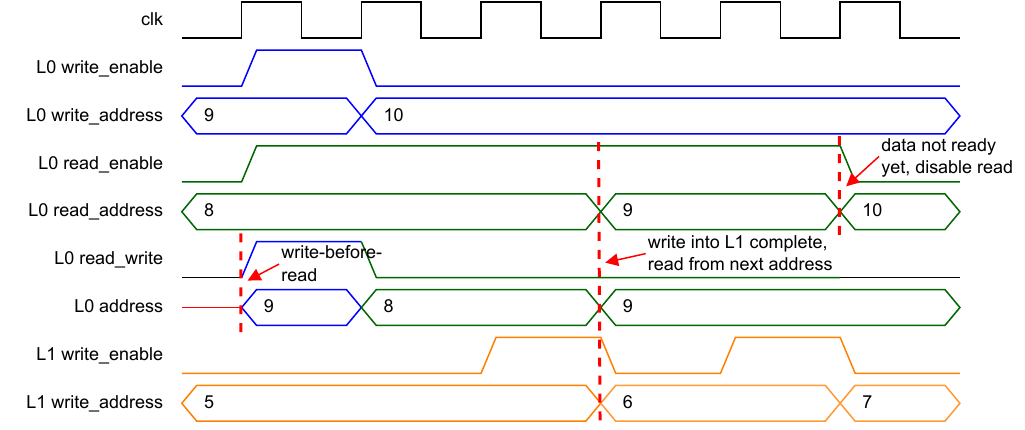}
	\caption[Synchronization of read and write in two hierarchy levels]{Waveform of read and write cycles of the two hierarchy levels L0 and L1. L0 is a single-ported module that prefers write over read accesses. \textit{L0 read\_write} toggles between read ($=0$) and write ($=1$) cycles. If L0 is in a write cycle, \textit{L0 write\_address} is forwarded to the address port of the memory module, otherwise \textit{L0 read\_address} is inserted. The reading of address 8 is postponed until the write into address 9 is complete. L1 is dual-ported so potential read requests can be ignored. The last read cycle at address 10 cannot be executed yet, since it is still waiting for data to be written into 10 first.}
  \label{fig:read_write}
\end{figure}

The \gls{mcu} manages the pattern state and controls memory accesses by using internal registers. 
These registers can be reset, facilitating the initiation of a new pattern.
Each hierarchy level, except the last, follows a consistent calculation process shown in Listing \ref{code:mcu}, with minor adjustments for level-specific settings.
Before a hierarchy level outputs data, it must be loaded from the input buffer or the preceding level (line $7-14$).
Internal pointers manage the writing address, ensuring systematic loading.
A pattern and an offset pointer are used to calculate the next address to be read from (l. $31$).
After each completed cycle, the pattern pointer will be reset and the offset pointer will be adjusted according to the selected inter-cycle shift (l. $20-28$).
As illustrated in Figure \ref{fig:read_write}, the \gls{mcu} only activates a write cycle, after the preceding level executed a read cycle.
A write-over-read policy prevents deadlocks of single-ported memories within the hierarchy.
Since a write needs an active read in the preceding level, the \gls{mcu} can at most activate the write mode every two clock cycles.
The last hierarchy level, positioned closest to the processing units, employs a dual-ported memory module for optimal performance.
The \textit{disable output} control signal stalls data output, allowing efficient data preloading and preventing premature outputs.

\subsubsection{Output Shift Register}
The \gls{osr}, an optional register file positioned between the memory hierarchy and the accelerator's processing units, boasts a fully configurable bit width that can surpass that of the last hierarchy level. This flexibility enables the \gls{osr} to store multiple words concurrently. Operating on each clock cycle, the \gls{osr} executes a left shift operation and, with sufficient register space, requests the next data word from the hierarchy. The \gls{osr}'s flexibility extends to configurable shifts during runtime, controlled by the \gls{mc}. However, it's essential to consider that shifts exceeding the bit width of the last hierarchy level can be impractical, as the memory module requires multiple cycles to fill the \gls{osr}, while the register could perform a shift every cycle. Furthermore, each additional available shift width contributes to increased chip size and energy consumption.

\section{Results}
\subsection{Verification}
The functional verification and simulation of the memory hierarchy written in SystemVerilog was realized using cocotb \cite{cocotb}. A Python model replicates the framework's functionality. Comprising an input buffer, multi-level storage hierarchy, and an \gls{osr}, the class utilizes the bitstring module \cite{griffiths2023bitstring} as back-end to efficiently support arbitrary word widths. The input buffer and \gls{osr} consist of individual bitstrings with configured bit width, while the memory hierarchy employs an array of bitstrings for each level, mirroring the combined RAM depth of all banks in a given level. During initialization, the model validates parameters, ensuring a valid configuration, and initializes components like the input buffer, memory hierarchy, \gls{osr}, and state variables. To accommodate different patterns or test cases, a new model instance can be initialized for each test, preventing interference from previous ones.

The model allows customization of the memory access pattern through an interface similar to the \gls{dut}. The pattern configuration is validated, catching invalid configurations that might lead to unknown system states. Pattern execution involves the two phases data insertion and retrieval. The model handles the off-chip data stream independently, transforming each new data word into a bitstring and storing it in the input buffer. Upon buffer fullness, the content is written into the array representing the first hierarchy level. Unused entries are promptly cleared to avoid data loss. The model's output and general behavior are updated selectively, reflecting only the design's external behavior.

\subsection{Performance Analysis}
\subsubsection{Increasing Cycle Lengths}
\begin{figure}[t]
	\centering
	\includegraphics[width=\linewidth]{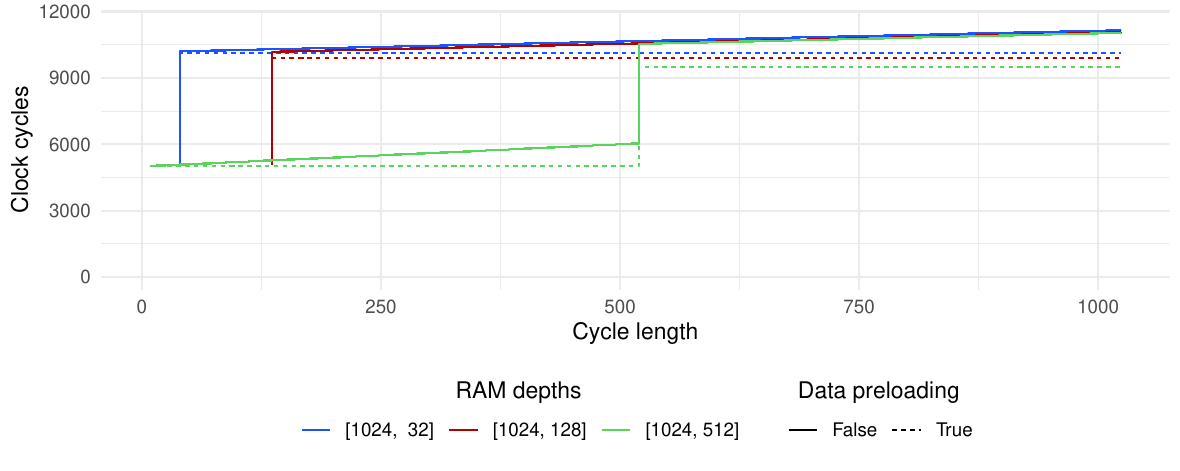}
	\caption[Cycle Lengths]{Impact on the required clock cycles to output 5,000 data words by increasing cycle lengths from 8 to 1,024 for the three given configurations with a \gls{ram} depth in level 1 of 32, 128, and 512, respectively. Each configuration was simulated with and without data preloading enabled.}
  \label{fig:test1}
\end{figure}
The memory framework aims to match the best-case performance of a standard \gls{sram} memory, targeting a data word output in every clock cycle. Three configurations were tested with increasing cycle lengths from 8 to 1,024, utilizing the same 1,024-word capacity for hierarchy level 0. However, level 1 had varying RAM depths: 32, 128, or 512 words. Tests were conducted both with and without data preloading. As shown in Figure \ref{fig:test1}, performance notably decreases after the cycle length surpasses the storage capacity of level 1, doubling the runtime due to internal data word replacement in a round-robin fashion and a write cycle limitation of every two cycles. Cycle lengths beyond level 1 capacity, larger memory hardly improves performance, with all hierarchies requiring similar clock cycles to complete the longest cyclic pattern. Preloading data enhances performance, as demonstrated by a 21\% decrease in clock cycles needed for the configuration with a 512 RAM depth level 1, emphasizing the potential for runtime optimization during pattern execution.

\subsubsection{Equal Capacity at Different Word Widths}
\begin{figure}[t]
	\centering
	\includegraphics[width=\linewidth]{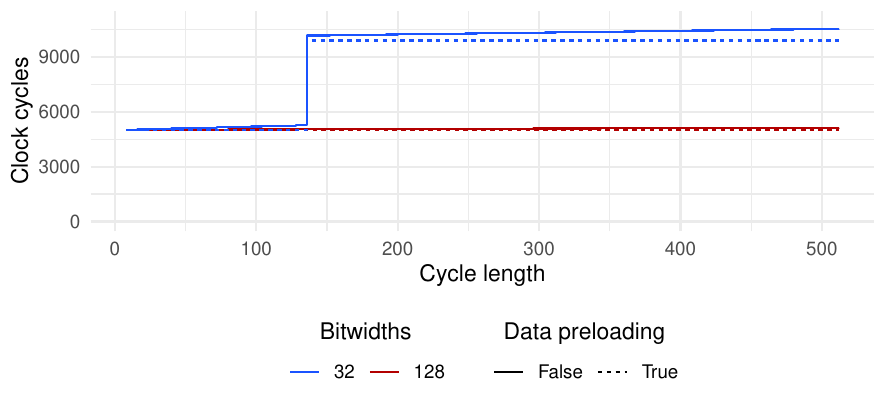}
	\caption{Influence on clock cycles needed to output 5,000 32-bit data words, spanning cycle lengths from 8 to 1,024, was examined with the bit widths 32- and 128-bit. The 128 bit configuration utilized an \gls{osr}. Each configuration was simulated with and without data preloading enabled.}
  \label{fig:test2}
\end{figure}

\begin{figure}[t]
    \centering
    \includegraphics[width=\linewidth]{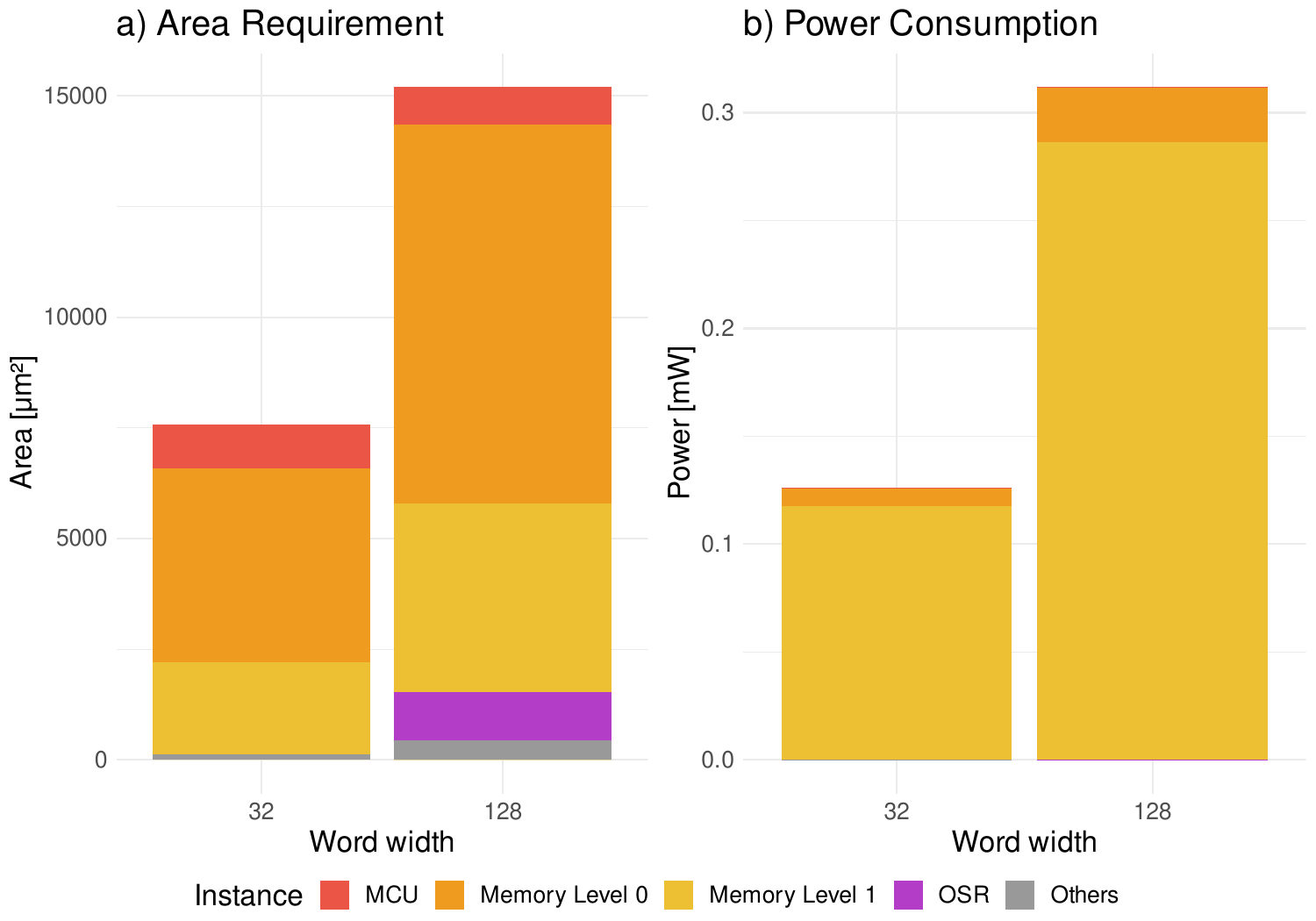}
    \caption[Chip area and power consumption of test 2 configurations]{Chip area and power consumption of both frameworks in Figure \ref{fig:test2}, which is dominated by the memory modules.}
    \label{fig:test2area_n_power}
\end{figure}
The second test case contrasts two memory frameworks, both having equal bit capacities. The first hierarchy, with a 32-bit word width, features RAM depths of 512 and 128 words for levels 0 and 1, respectively. In contrast, the second hierarchy adopts a 128-bit word width with shallower RAM depths of 128 and 32 words for the same levels and includes an \gls{osr} to output 32-bit words. Figure \ref{fig:test2} depicts the clock cycles required to generate 5,000 outputs across varying cycle lengths. The first hierarchy mirrors the previous test case, doubling its clock cycles beyond a cycle length of 128 words. However, the second hierarchy, with a wider word width, consistently performs optimally throughout all cycle lengths, copying four 32-bit words per write cycle. Although this wider word width significantly increases hierarchy costs, as evident in Figure \ref{fig:test2area_n_power}, doubling the required chip area from \SI{7566}{\micro\metre^2} to \SI{15202}{\micro\metre^2}, it also incurs a higher energy consumption of \SI{0.31}{\milli\watt}, nearly 2.5 times more than the 32-bit architecture.

\subsubsection{Impact of the Inter-Cycle Shifts}
\begin{figure}[t]
	\centering
	\includegraphics[width=\linewidth]{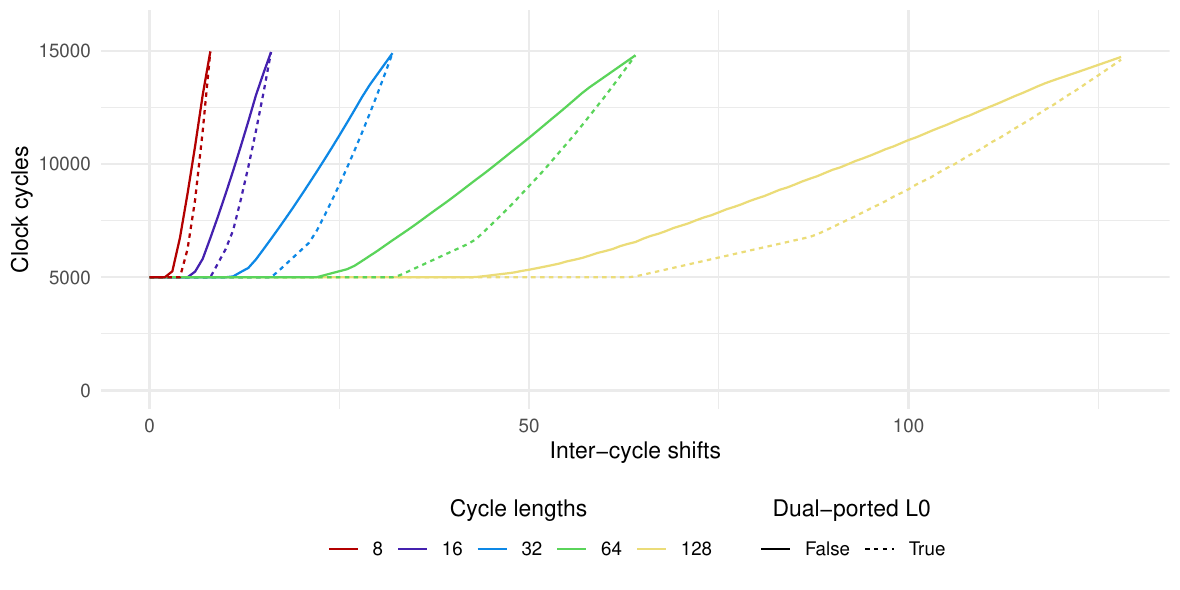}
	\caption[Increasing inter-cycle shifts]{Impact on the required clock cycles to output 5,000 data words by increasing inter-cycle shifts at selected cycle lengths. The hierarchies are configured with two levels and \gls{ram} depths of 512 and 128 words. One hierarchy has a single-ported L0 memory while the second configuration has a dual-ported module. The inter-cycle shift was configured from 1 to the currently selected cycle length.}
  \label{fig:test3}
\end{figure}
The performance impact of the inter-cycle shift on the memory framework was evaluated using configurations with two hierarchy levels. Both levels had a 32-bit word width, with level 0 and level 1 having RAM depths of 512 and 128 words, respectively. One configuration utilized a single-ported memory for level 0, while the second configuration employed a dual-ported module. Despite sharing the same pattern selections, the test revealed notable differences in throughput. Figure \ref{fig:test3} displays results for increasing inter-cycle shifts at static cycle lengths, indicating optimal throughput when the inter-cycle shift is less than one-third of the cycle length. However, performance decreases significantly with larger inter-cycle shifts, reaching the worst-case scenario with an output every three clock cycles when the inter-cycle shift equals the cycle length. The dual-ported design delays this performance decline but doesn't improve the worst-case scenario. In cases requiring data reloading from off-chip memory, the bottleneck is the single-ported memory module in level 0, affecting throughput and synchronization with the input buffer. The dual-ported level 0 enhances performance at larger inter-cycle shifts with minimal additional chip area, as depicted in Figure \ref{fig:test3}. However, the power consumption increases by 130\%, emphasizing the need to balance performance gains with energy requirements in the final design.

\subsection{Case Study}
\begin{table*}[t]
    \small
    \begin{tabularx}{\textwidth}{ l *{13}{Y} }
    \hline
    \rowcolor[HTML]{EFEFEF} 
    Layer & 0 & 1 & 2 & 3 & 4 & 5 & 6 & 7 & 8 & 9 & 10 & 11 & 12 \\ \hline
    \cellcolor[HTML]{EFEFEF}\begin{tabular}[c]{@{}l@{}}Layer\\ Type\end{tabular} & CONV & CONV & CONV & CONV & CONV & CONV & CONV & CONV & FC & CONV & CONV & CONV & FC \\
    \cellcolor[HTML]{EFEFEF}\begin{tabular}[c]{@{}l@{}}Unique\\ Addresses\end{tabular} & 1,920 & 3,456 & 384 & 5,184 & 6,912 & 768 & 9,216 & 512 & 196 & 13,824 & 1,536 & 20,736 & 768 \\
    \cellcolor[HTML]{EFEFEF}\begin{tabular}[c]{@{}l@{}}Cycle\\ Length\end{tabular} & 98 & 45 & 49 & 41 & 20 & 24 & 16 & 24 & 1 & 8 & 12 & 4 & 1 \\ \hline
    \end{tabularx}
    \caption{Type, number of unique addresses, and cycle lengths of the shifted cyclic pattern of each TC-ResNet layer.}
    \label{tab:layer}
\end{table*}
To analyze the capabilities of the memory framework, a case study was performed with the low power hardware accelerator UltraTrail \cite{bernardo2020ultratrail}.
It is trained for \gls{kws} to classify a subset of the Google speech commands data set and utilized a TC-ResNet model.
The accelerator is standardly equipped with 64 \gls{mac} units arranged in an $8 \times 8$ array.
Therefore, it is desirable to unroll the loop nests of the \gls{dnn} layers in a fashion to maximize the utilization of the processing units.
The memory access patterns of the weight and input data sets of all feasible unrollings were analyzed.

Each convolutional layer can be unrolled along its factors batch size $N$, number of groups $G$, number of output channels $K$, number of input channels $C$, width of input channel $X$, and filter width $F$.
However, since the accelerator's data flow is mostly static, each layer must be unrolled along the same factors.
The resulting memory traces of the selected unrolling can be analyzed to determine performance predictions.
To optimize the runtime needed to execute the \gls{dnn}, a high average data-parallelism should be possible within all layers with the given unrolling.
Otherwise, the utilization of the \gls{mac} array will be low.
Additionally, the number of unique data words per loop step needs to be considered as they dictate the required port width of the data set.
Wide ports complicate routing of the chip and possibly require multiple memory banks to support the word widths.
Finally, the memory access pattern can be evaluated using the memory traces.
Complex nested patterns with a low data reuse rate should be avoided as they need to utilize a larger memory hierarchy to support their difficult memory accesses.

The weight data sets exhibit a sequential pattern, allowing a single-level, small memory hierarchy to efficiently handle the access pattern, minimizing capacity needs and chip area. In contrast, the input data sets, with more intricate patterns depending on loop unrolling and layer, may require multiple levels. Some unrolling scenarios currently lack \gls{mcu} support, necessitating the storage of complete nested patterns, significantly increasing capacity requirements. Under worst case conditions, data from a single off-chip address must be stored several times in the hierarchy, highlighting the challenge of efficiently managing complex input patterns. If parallel access patterns were supported, a more optimized memory configuration could be employed, potentially reducing chip area and power consumption.

\subsubsection{Memory Hierarchy compared to Dual-ported SRAMs}
Originally designed for 8-bit inputs and 6-bit weights, in this subsection we suppose that UltraTrail now employs 8-bit data words and a 32-bit off-chip memory architecture for both data sets, simplifying memory module selection.
In this context, a comprehensive comparison emerges between a manually configured memory framework and basic dual-ported modules, aimed at achieving optimal throughput.
Significantly, the memory hierarchy and dual-ported memories align in word width.
Since the weight dataset always has a simple shifted cyclic access pattern, the framework was configured with only one hierarchy level.
A multi level framework would increase chip area and power consumption, while the offered pattern space would not be utilized.

\begin{figure}[t]
	\centering
	\includegraphics[width=\linewidth]{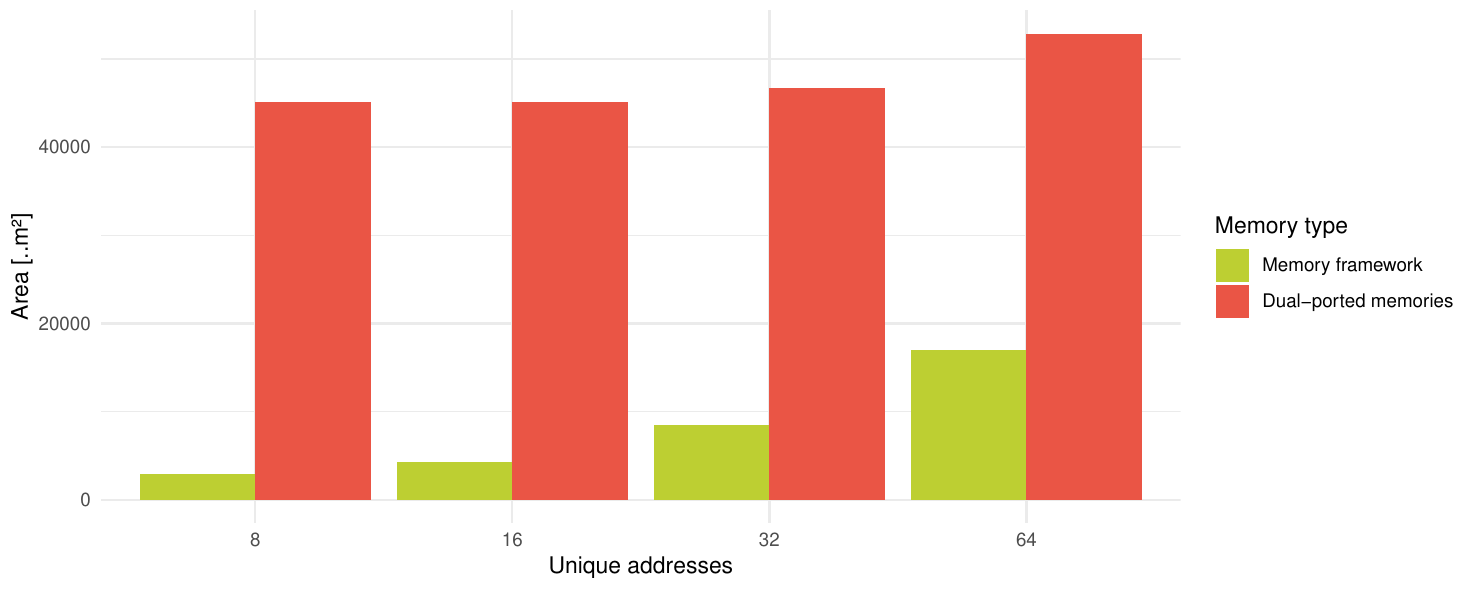}
	\caption{Comparison of the occupied chip area between dual-ported memory modules with enough capacity to store all data words and the memory frameworks that can perform the access pattern of all layers.}
  \label{fig:weights_area}
\end{figure}
\begin{figure}[t]
	\centering
	\includegraphics[width=\linewidth]{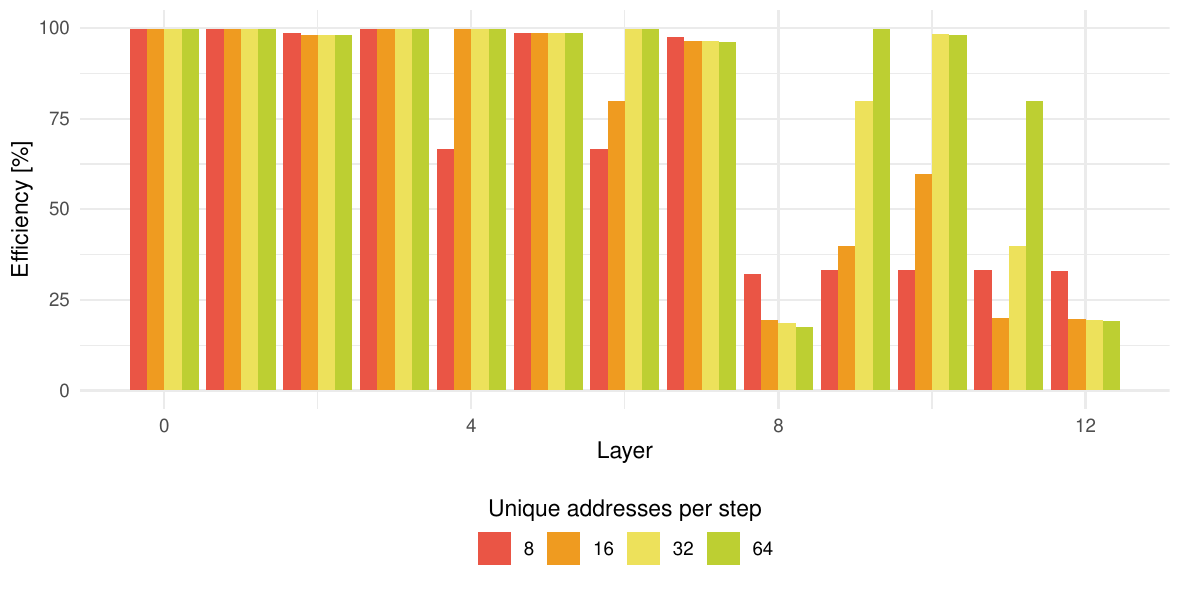}
	\caption{Relative runtime of each TC-ResNet layer with the memory framework using unrollings with 8, 16, 32, and 64 unique addresses per step. The theoretical optimal runtime would be one output in each cycle.}
  \label{fig:weights}
\end{figure}
Regarding the weights, layer eleven of the TC-ResNet model has the highest capacity requirement among all layers with 20,736 unique data words.
Unrollings featuring only eight unique addresses per loop step, demand a 64-bit word width with at least 2,592 RAM depth.
While dual-ported 64-bit memory can only offer a maximum capacity of 2,048, necessitating two banks, the memory framework efficiently uses a single 64-bit dual-ported memory with a capacity of 32 words, occupying only 6.5\% of the chip area compared to the dual-ported alternatives.
For unrollings with 16 distinct addresses, either two 128-bit banks (accessed consecutively) or two 64-bit banks (working in parallel) are required.
However, unrollings with 32 and 64 unique addresses need multiple banks for data parallelism.
As shown in Figure \ref{fig:weights_area}, despite a 17.1\% increase, the dual-ported \glspl{sram} remain 3.1 times larger than the parallel memory frameworks.
Figure \ref{fig:weights} illustrates runtime comparisons, demonstrating the memory framework's relative efficiency of 58.8\%, 60.6\%, 85.7\%, and 97.6\% for 8, 16, 32, and 64 unique addresses per step, respectively.
Note that an efficiency of $100 \%$ represents one data word output in each clock cycle.
However, the framework was executed without any preloading, which resulted in a loss of performance due to the need for data fetching.
This highlights the potential for further improvements by using idle time between layers for data preloading.

\begin{figure*}[t]
    \centering
    \includegraphics[width=\textwidth]{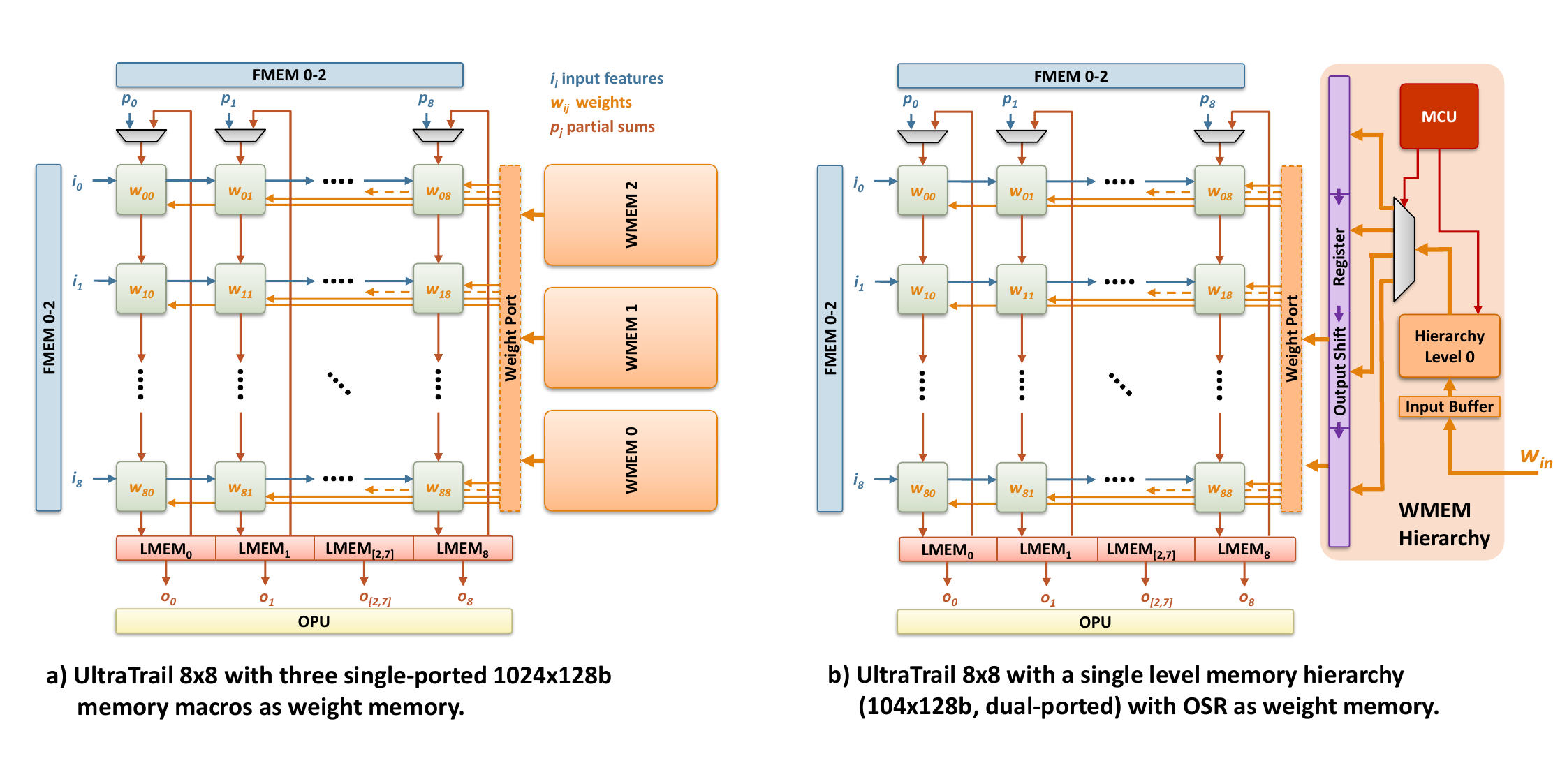}
    \caption{Two UltraTrail 8x8 configurations adapted from \cite{bernardo2020ultratrail}. Each of the 64 \gls{mac} units needs a 6-bit weight, resulting in a 384-bit weight port width.}
    \label{fig:utwithmemfw}
\end{figure*}
\begin{figure}[t]
	\centering
	\includegraphics[width=\linewidth]{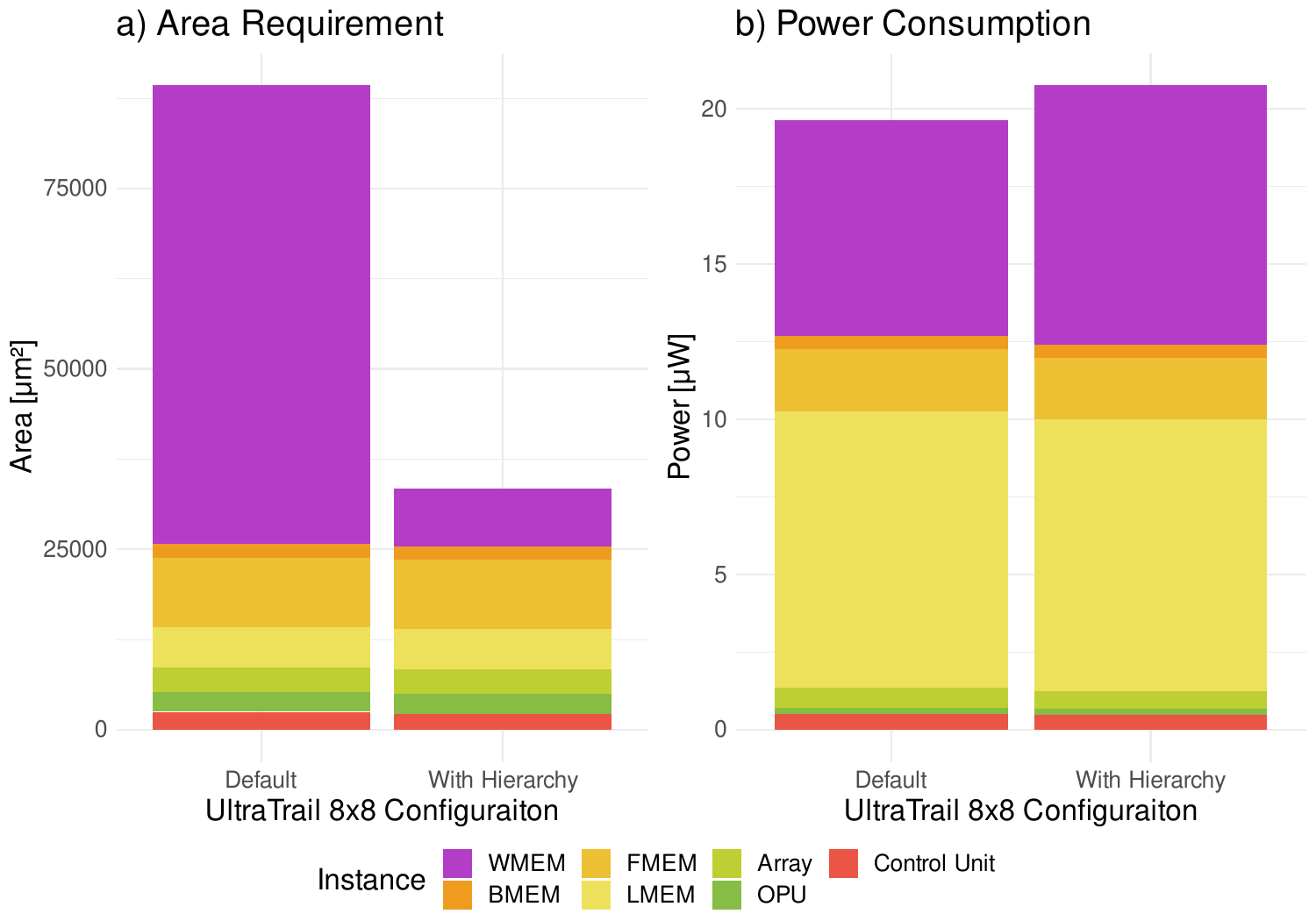}
	\caption{Syntheses results of chip area and power consumption of a baseline UltraTrail 8x8 and a configuration with the memory hierarchy as WMEM.}
  \label{fig:integration}
\end{figure}

Shifting focus to input data sets, the limitations of the memory framework become evident, as certain patterns are not supported efficiently yet.
An updated \gls{mcu} is needed to address these complex parallel nested patterns to achieve optimal performance.

\subsubsection{Integration as Weight Memory}
The standard UltraTrail 8x8 configuration utilizes three single-ported $1024$x$128$-bit \gls{sram} modules to be able to store the complete weight data set of the TC-ResNet model as shown in Figure \ref{fig:utwithmemfw}a. These macros alone occupy more than $70\%$ of the accelerators chip area.
After we replaced the memory modules with a single level memory hierarchy, the accelerators chip area was reduced by $62.2\%$. The hierarchy has one dual-ported memory module with a 128-bit word width. An \gls{osr} is used to generate the required word with of 384 bits. 

The hierarchy needs three clock cycles to fill the \gls{osr} if the data is already stored in the memory level 0.
The time needed loading these weights from the off-chip memory depends on the three factors off-chip word width, off-chip clock frequency, and average response time.
Our test bench simulated an 32-bit off-chip architecture that is clocked at \SI{1}{\mega\hertz}.
The accelerator is clocked four times slower at \SI{250}{\kilo\hertz}.
This very slow clock rate was chosen to meet the real-time requirement of \SI{100}{\milli\second} per inference while reducing the power consumption of the chip \cite{bernardo2020ultratrail}.
Data requests of the hierarchy were delivered with a latency of one clock cycle.
As a result, three accelerator clock cycles were needed to request and store a 128-bit weight within the hierarchy.

The loop-nest analysis of the TC-ResNet revealed that only \gls{fc} layers (layer eight and twelve) do not reuse their weights for multiple loop iterations.
As exposed by Table \ref{tab:layer}, these layers do not dominate the computational costs of the neural network.
Thus, their low efficiency can be ignored.
At four, the \gls{conv} layer eleven has the shortest cycle length.
These cycle lengths are too short for the hierarchy to be able to read data from the off-chip memory and provide it to the \gls{mac} array in time.
However, these delays could be reduced by preloading the data during the execution of preceding layers.
\newpage
Three inference runs were used to estimate the power consumption of the adjusted UltraTrail and to compare it to its baseline counterpart. As illustrated in Figure \ref{fig:integration}b, the power consumption does slightly increase by $6.2\%$ if the hierarchy is used instead of standard single-ported modules. This is due to the significantly greater leakage power of dual-ported memory compared to single-ported memories.

\subsection{Discussion and Future Work}
The case study revealed the potential of an on-demand streaming interface for hardware accelerators.
Instead of loading the weight data set once and storing it in large on-chip memory macros before the first inference, the hierarchy enables a significant reduction of required chip area.
At the same time, the hierarchy increases the accelerator's flexibility by enabling it to switch between different \glspl{dnn} more frequently.
To start the execution of another \gls{dnn}, the hierarchy just needs a reset cycle with the new pattern settings.
However, the power consumption does slightly increase as dual-ported memories have higher leakage and switching power than their single-ported counterparts.
Also, the continuous off-chip memory accesses result in additional power consumption.
The actual amount of energy consumed, however, does highly depend on the \gls{soc} and its memory system.
A slight redesign with a dual-banked, single-ported hierarchy could solve this issue with only a minor chip area overhead.

The framework achieved promising results regarding the weight data sets, however, it should be improved to be efficiently utilizable for the input data set.
A more modular and customizable design could increase the supported pattern range.
Additionally, the redesign could also further increase the throughput of the hierarchy, if required.

\section{Conclusion}
This paper presented a configurable memory hierarchy framework dedicated for neural network hardware accelerators. By exploiting the calculability of \gls{dnn} memory access patterns, it is possible to deploy smaller memory configurations than the conventional approach of storing the complete data set on the accelerator. The on-demand data fetching and delivery increases the possibilities of the accelerator's design space. The performance analysis verified the efficient support of common access patterns. Finally, the UltraTrail case study showed, that the hierarchy can reduce the required chip area by up to $62.2\%$ while only resulting in a performance degradation of $2.4\%$.

\bibliographystyle{IEEEtran}
\small
\balance
\bibliography{bibliography.bib}
\end{document}